\documentclass[structabstract]{aa}  
\usepackage{graphicx}
\usepackage{natbib}
\usepackage{longtable,lscape}
\usepackage{txfonts}
\begin{document}
   \title{XMM-{\it Newton} detection of the supernova remnant G304.6$+$0.1 \\(Kes 17)}
\author{J.~A. Combi\inst{1,3,4}, J.F. Albacete Colombo\inst{2}, E. S\'anchez-Ayaso\inst{3}, G.E. Romero\inst{1,4}, J. Mart\'{\i}\inst{3}, P.L. Luque-Escamilla\inst{5}, A.J. Mu\~noz-Arjonilla\inst{3}, J.R. S\'anchez-Sutil\inst{3}, J. L\'opez-Santiago\inst{6}
}
\authorrunning{Combi et~al.}
\titlerunning{Detection of X-ray emission of the SNR G304.6$+$0.1} 

\offprints{J.A. Combi}  

   \institute{Instituto Argentino de Radioastronom\'{\i}a (CCT La Plata, CONICET), C.C.5, (1894) Villa Elisa, Buenos Aires, Argentina.\\
              \email{[jcombi:romero]@fcaglp.unlp.edu.ar}
         \and
             Centro Universitario Regional Zona Atl\'antica (CURZA). Universidad Nacional del COMAHUE, Monse\~nor Esandi y Ayacucho (8500), 
Viedma (Rio Negro), Argentina.\\
             \email{donfaca@gmail.com}
         \and
         Departamento de F\'{\i}sica (EPS), Universidad de Ja\'en, Campus Las Lagunillas s/n, A3, 23071 Ja\'en, Spain\\
\email{[esayaso:jmarti:ajmunoz:jrssutil]@ujaen.es}
\and
Facultad de Ciencias Astron\'omicas y Geof\'{\i}sicas, Universidad Nacional de La Plata, Paseo del Bosque, B1900FWA La Plata, Argentina.
\and
Departamento de Ingenier\'{\i}a Mec\'anica y Minera, Escuela Polit\'ecnica Superior, Universidad de Ja\'en, Campus Las 
Lagunillas s/n, A3, 23071 Ja\'en (Spain).\\
\email{peter@ujaen.es} 
\and
Departamento de Astrof\'{\i}sica y Ciencias de la Atm\'osfera, Universidad Complutense de Madrid, E-28040, Madrid, Spain.\\
\email{jls@astrax.fis.ucm.es}  
             }

   \date{Received ; accepted }

 
  \abstract
   {}
   {We report the first detailed X-ray study of the supernova remnant (SNR) G304.6$+$0.1, achieved with the XMM-{\it Newton} mission.}
   {The powerful imaging capability of XMM-{\it Newton} was used to study the X-ray characteristics of the remnant at different energy ranges. The X-ray morphology and spectral properties were analyzed. In addittion, radio and mid-infrared data obtained with the Molonglo Observatory Synthesis Telescope and the Spitzer Space Telescope were used to study the association with the detected X-ray emission and to understand the structure of the SNR at differents wavelengths.}
   {The SNR shows an extended and arc-like internal structure in the X-ray band with out a compact point-like source inside the remnant. We
find a high column density of $N_{\rm H}$ in the range 2.5-3.5$\times$10$^{22}$ cm$^{-2}$, which supports a relatively distant location ($d \geq$ 9.7 kpc). The X-ray spectrum exhibits at least three emission lines, indicating that the X-ray emission has a thin thermal plasma origin, although a  non-thermal contribution cannot be discarded. The spectra of three different regions (north, center and south) are well represented by a combination of  a non-equilibrium ionization (PSHOCK) and a power-law (PL) model. The mid-infrared observations show a bright filamentary structure along the north-south direction coincident with the NW radio shell. This suggests that Kes 17 is propagating in a non-uniform environment with high density and that the shock front is interacting with several adjacent massive molecular clouds. The good correspondence of radio and mid-infrared emissions suggests that the filamentary features are caused by shock compression. The X-ray characteristics and well-known radio parameters indicate that G304.6$+$0.1 is a middle-aged SNR (2.8-6.4)$\times$10$^{4}$ yr old and a new member of the recently proposed group of mixed-morphology SNRs.}
   {}


   \maketitle
%

\section{Introduction}

\begin{figure*}
\centering
\includegraphics[width=18.2cm]{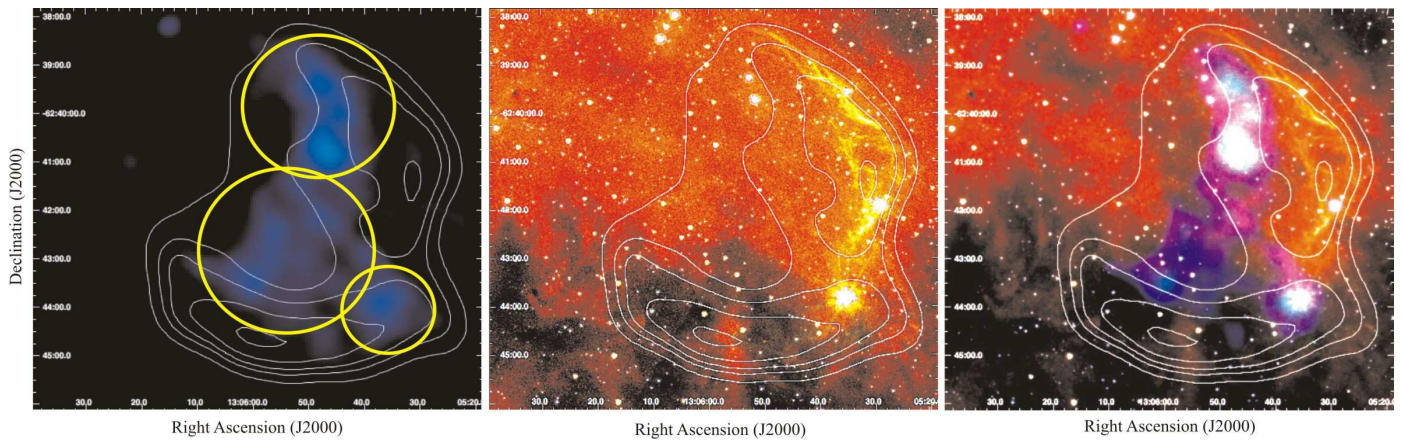}
\caption{{\bf Left panel:} XMM-{\it Newton} color image of the X-ray emission in the 0.3-10 keV energy range (in blue) of G304.6$+$0.1 with the radio contours (in white) at 843 MHz overlaid. The extraction regions used for the spectral analysis are also indicated as yellow circles. {\bf Central panel:} IRAC three-channels composed image of G304.6$+$0.1 with the radio contours (in white) at 843 MHz overlaid. {\bf Right panel:} Composite color image of the radio, mid-infrared, and X-ray emissions of G304.6$+$0.1. All images are in equatorial coordinates J2000.}
\label{FigVibStab}
\end{figure*}

The supernova remnant G304.6$+$0.1 (also known as Kes 17) was first detected at 408 MHz and 5 GHz by Shaver \& Goss (1970), using the Parkes and Molonglo radiotelescopes. The authors obtained spectral information and some physical characteristics of the SNR. A spectral index $\alpha \sim -0.51$ (S $\propto \nu^{\alpha}$), an angular size of 8.4 arcmin and a possible distance of 12 kpc were estimated for the source. Afterwards, Milne \& Dickel (1975) determined a radio  polarization of 1.3\%. Using the Parkes hydrogen line interferometer, Caswell et al. (1975a) constrained the distance to a minimum of 9.7 kpc. Later, Milne et al. (1985) mapped the object with the Fleurs array, and suggested that its peculiar structure resulted from the shock front collision with a dense insterstellar cloud. The source was then reported in the Molonglo Observatory Synthesis Telescope (MOST) Catalog (Whiteoak \& Green 1996) as a small-diameter SNR with irregular shell morphology. In addittion, Frail et al. (1996) have detected 1720 MHz OH maser  emission around the object using the Green Bank telescope. 

More recently, Reach et al. (2006), using mid-infrared data from the Spitzer Space Telescope (Werner et al. 2004) at 3.6, 4.5, 5.8, and 8$\mu$m of G304.6$+$0.1, revealed bright thin isolated filaments in each of the IRAC channels. Based on colors and the detailed morphological agreement of the images in the four channels, the authors suggested that most of the mid-infrared emission from the shell could be produced in the shocked molecular material. Recently, Hewitt et al. (2009) computed the total hydrogen column density along the line of sight to the SNR, using X-ray observations of archival ASCA data and found diffuse interior X-ray emission. However, at present no X-ray study of the source has been published.

We report an X-ray study of SNR G304.6$+$0.1, using the greatly enhanced sensitivity of the XMM-{\it Newton} mission. An instrument with these characteristics offers a unique opportunity to detect X-ray emission from distant and obscured SNRs. Radio and infrared observations were also used for a multiwavelength study of the source. This work is part of a program aimed to identify and study X-ray emission from SNRs. The structure of the paper is as follows: in Sect.~\ref{observations} we describe the XMM-{\it Newton} observations and data reduction. X-ray analysis and radio/infrared results are presented in Sects 3 and 4, respectively. In Sect.~\ref{discusion}, we discuss the implications of our results. Finally, we summarize the main conclusions in Sect.6. 

\section{Observations and data reduction} \label{observations}

The field of SNR G304.6+0.1 was observed on 2005 August 25 
by the XMM-{\it Newton} X-ray satellite during revolution 1046 (Obs-Id. 0303100201), 
with a total integration time of 26.8 ksec. 
The observation was centered on 
$(\alpha, \delta)_{\rm J2000.0} = (13^{\rm h}05^{\rm m} 46\fs0, -62\degr 43\arcmin 40\arcsec)$, 
and acquired by the 
EPIC-MOS (Turner et al., 2001) and EPIC-PN (Struder et al. 2001). 
The data were taken with a medium filter in full-frame (FF) imaging mode. 
The data were analyzed with the XMM Science Analysis System (SAS) 
version 9.0.0 and the latest calibrations.

To exclude high background flares, which could eventually affect the observations, 
we extracted light curves of photons above 10 keV for the entire 
field-of-view of the EPIC, and discarded time intervals in
which background flares occurred for subsequent analysis.
We finally used a total of 21.181 ksec, 23.962 ksec, and 23.957 ksec for the
PN, MOS-1 and MOS-2, respectively. To create images, spectra, and light curves, we selected events with
FLAG=0, and PATTERNS= 12 and 4 for MOS and PN cameras, respectively. 
Hereafter we use clean event-files in the [0.3-10] keV energy band.

\section{X-ray analysis of G304.6$+$0.1} \label{analysis}
\subsection{Image}

In Fig.1 (left panel) we show the X-ray image of G304.6$+$0.1 in the 0.5-0.8 keV energy range. 
The total unabsorbed X-ray flux for this band (after the fit to the X-ray spectrum, see Table 1 for details) is $F_{\rm X (0.5-8.0~keV)}$= 
7.11($\pm 0.36$) $\times 10^{-11}$ erg~cm$^{-2}$~s$^{-1}$.

The image reveals diffuse and extended X-ray emission with an irregular filled-center and possible arc-like features. No compact point-like X-ray source was found within the diffuse emission. Assuming a central source with a typical compact central object (CCO) spectrum ($\Gamma \sim$ 5), we obtain a flux upper limit of 6.5$\times$10$^{-15}$ erg~cm$^{-2}$~s$^{-1}$ in the 0.5-8 keV band. Clearly the X-ray emission seems to correlate well with the internal part of the NW radio shell. An irregular region of X-ray emission is also observed at $(\alpha, \delta)_{\rm J2000.0} = (13^{\rm h}05^{\rm m} 35\fs0, -62\degr 44\arcmin 00\arcsec)$.

\begin{figure}
\centering
\includegraphics[width=8.5cm,angle=0]{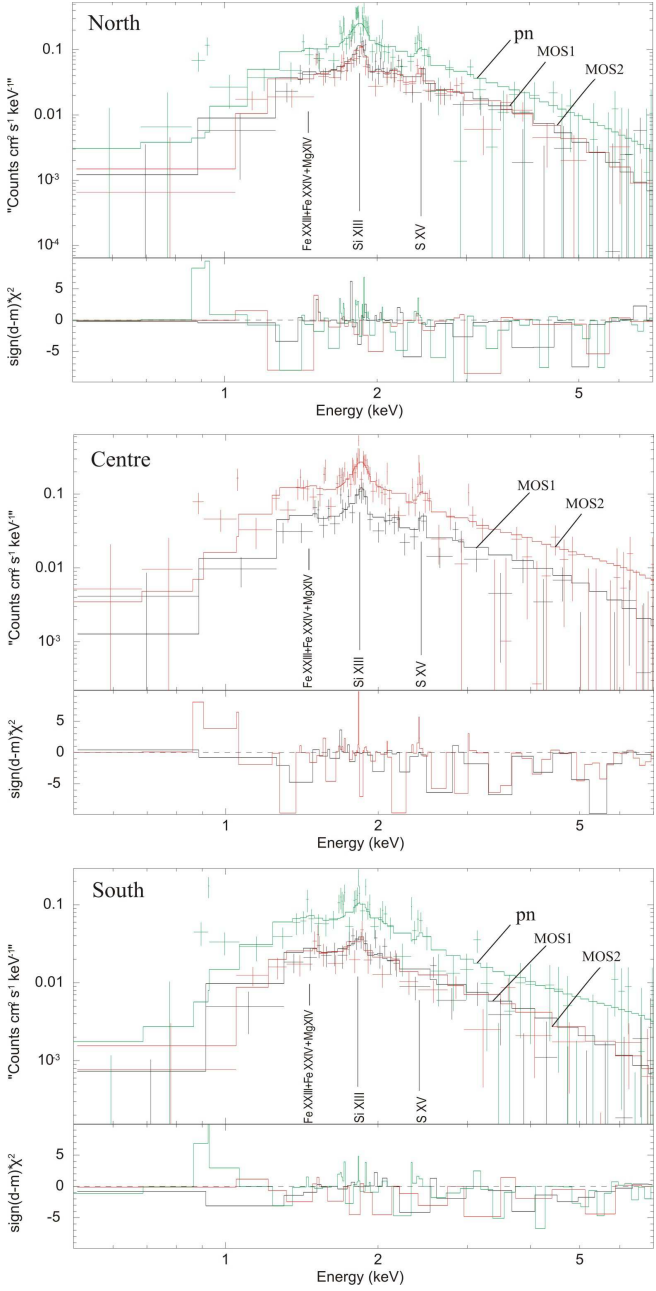}
   \caption{Combined XMM-{\it Newton} EPIC {\sc PN} and {\sc MOS} X-ray spectra for the regions north, center, and south of G304.6$+$0.1 and residuals (botton panel in each spectrum).
   The solid line indicates the best-fit {\sc pshock}+PL model (see Table 1).
   The identified lines, mostly blended, correspond to atomic transitions of
   Fe XXIII+Fe XXIV+Mg XIV, Si XIII, and S XV.
}
\end{figure}

\subsection{Spectral analysis} 

The X-ray spectrum of G304.6+0.1 was extracted from the EPIC cameras.
We used the SAS task \textsc{evselect} with appropiate parameters for the PN and MOS\,1/2
cameras. The extraction regions used were circles with radii 1.4, 1.8, and 0.75 arcmin (see Fig.1, left panel).
Background spectra were extracted from a region where no X--ray emission was detected.

Figure 2 shows the background-substracted EPIC-PN spectra obtained from the XMM-{\it Newton}
observation for three differents regions (i.e., north, center, and south). We grouped the extracted EPIC PN and MOS
spectra with a minimum of 25 and 16 counts per spectral bin.
Ancillary response files (ARFs) and redistribution matrix files (RMFs)
were calculated. At high X-ray energies (above 7 keV), the full spectrum shows features
with low statistical significance (1 to 1.5 $\sigma$), which are probably
related to fluorescence lines in the background spectrum of 
XMM-{\it Newton} (e.g. De Luca \& Molendi, 2004). The spectral analysis
was performed with the XSPEC package (Arnaud, 1996). The X-ray spectrum of the remnant exhibits emission lines 
at energies of 1.54 keV (Fe XXIII+Fe XXIV+Mg XIV),
1.85 keV (Si XIII), and 2.45 keV (S XV). However, the quality of the data prevents a 
further refined analysis of the abundances of each single ion.\\

The X-ray spectrum was fitted with various models: ($i$) a {\sc pshock} non-equilibrium ionization model with constant temperature (Mazzotta et al.1998); ($ii$) a {\sc mekal} emission model from hot diffuse gas (Mewe et al. 1985), and ($iii$) a single power-law (PL) model. Each model was  modified by a low-energy absorption model WABS (Morrison \& McCammon 1983). Several combinations were tentatively used to fit the spectra in these regions. We initially varied individual parameters: absorption ($N_{\rm H}$), temperature ($kT$), $\Gamma$ index,
ionization time scale ($\tau$), and normalization during the minimization process. As a first step, we tried the spectral fit 
by using single thermal models, but they did not yield acceptable fits at hard energies (i.e. $E >$ 4 keV), leading to large residual statistics, i.e.
$\chi_\nu^2 \sim$ 2.0 or even larger. Otherwise, the non-thermal emission model failed to fit the spectrum 
at the soft energies, affecting the $N_{\rm H}$ determination. We decided to treat the abundance as a single parameter in the {\sc pshock} model. 
We initially used the solar abundances of Anders \& Grevesse (1989), which were also
thawed during the minimization procedure. The goodness of the model fit was derived according to
the 2-test statistics.\footnote{C-statistics was also applied to the spectral
fit with similar results.}

A combination of two thermal models works well, but the temperature of the second
(hard) component is mismatched, therefore it does not provide a satisfactory description of the
X-ray spectrum. It reaches temperatures higher than 10 keV, which
cannot be considered a reasonable solution for a restricted 0.3-8.0 keV X--ray spectrum.
The results suggest that the thermal emission dominates at low and intermediate energies,
while a small contribution of non-thermal emission ($\sim$8$\%$) is important at high energies. 
Therefore, on the basis of $\chi^2$ minimization, a combination of thermal and non-thermal models provides the best acceptable fit description of the X--ray spectra of the of G344.6+0.1 in the three regions. The power-law model adequately fits 
high-energy photons (up to 5 keV) in which particle acceleration in shock-fronts could physically 
explain this X--ray emission. The parameter values of the thermal and non-thermal models are physically reasonable (see Table 1 for details).
Unfortunately, the sensitivity and spatial/spectral resolution is not suitable to isolate the thermal and non-themal X-ray contributions.

\begin{table*}
\caption{X--ray spectral parameters of G344.6+0.1}
\label{spec}
\begin{center}
\begin{tabular}{lccc}
\hline
Model & 
\multicolumn{1}{c}{South}  & 
\multicolumn{1}{c}{Center} & 
\multicolumn{1}{c}{North} \\
Parameters & & & \\
\hline
{\bf WABS} & & \\
N$_{\rm H}$ [cm$^{-2}$]&2.5($\pm$0.2)$\times$10$^{22}$ & 3.2($\pm$0.2)$\times$10$^{22}$& 3.5($\pm$0.3)$\times$10$^{22}$ \\
\hline
{\bf PSHOCK} & & & \\
kT [keV] &0.85($\pm$0.1) &0.72($\pm$0.1) & 0.68($\pm$0.09)\\
Abundance  &0.3($\pm$ 0.1) & 0.9($\pm$0.3)&1.2($\pm$0.6)\\
$\tau$[s cm$^{-3}$]  &3.1($\pm$0.7)$\times$10$^{12}$ &2.0($\pm$0.6)$\times$10$^{12}$ & 1.2($\pm$0.7)$\times$10$^{12}$\\
Norm &9.0($\pm$0.2)$\times$10$^{-4}$ &1.3($\pm$0.1)$\times$10$^{-3}$ 
& 6.0($\pm$0.2)$\times$10$^{-3}$\\
EM ($\times$10$^{57}$) &  5.8327562 &  12.142146 & 9.1539998     \\
{\bf POWER LAW} & & &\\
$\Gamma$ & 2.4($\pm$ 0.6) &1.8($\pm$ 0.7) & 3.1($\pm$ 0.3) \\
Norm  &1.1($\pm$0.9)$\times$10$^{-4}$&2.0($\pm$0.7)$\times$10$^{-4}$ &6.1($\pm$0.7)$\times$10$^{-4}$\\
\hline
$\chi^2$ / d.o.f. &0.76 / 2030  &0.83 / 3016 &0.8 / 2131\\
Flux[erg~cm$^{-2}$~s$^{-1}$]  &0.5($\pm$0.2)$\times$10$^{-11}$ &3.18($\pm$0.07)$\times$10$^{-11}$ & 3.43($\pm$0.09)$\times$10$^{-11}$  \\
\hline
\end{tabular}
\end{center}
Normalization is defined as 10$^{-14}$/4$\pi$D$^2$$\times \int n_H\,n_e dV$,
where $D$ is distance in [cm], n$_{\sc H}$ is the hydrogen density [cm $^{-3}$], $n_e$
is the electron density [cm$^{-3}$], and $V$ is the volume [cm$^{3}$]. The flux is 
absorption-corrected in the 0.5-8.0 keV energy range.
Values in parentheses are the single parameter 90\% confidence interval.
The abundance parameter is given relative to the solar values of Anders \& Grevesse (1989).
\end{table*} 

\section{Radio and infrared characteristics of G304.6$+$0.1}

To investigate the correlation between the X-ray, radio, and mid-infrared emissions of G304.6$+$0.1, we used radio observations performed at 0.843 GHz (with a resolution of 43") by the MOST, and mid-infrared data of the Infrared Array Camera (IRAC) (Fazio et al. 2004), obtained by the Spitzer Space Telescope at 3.6, 4.5, 5.8, and 8$\mu$m, respectively. Figure 1 shows three images that allow us to compare the correlation among the radio, infrared,  and X-ray emissions. Figure 1, left panel, shows the X-ray emission of the SNR in the 0.3-10 keV energy range, with the 843 MHz radio contours (in white) overlaid. Figure 1, central panel, shows the Spitzer/IRAC mid-infrared color image of the region where G304.6$+$0.1 lies, with the radio contours at 843 MHz (in white) superimposed. Finally, a composite color image of the radio/mid-infrared and X-ray emission of G304.6$+$0.1 is shown in Fig.1, right panel.

At radio frequencies, the source displays an asymmetric bilateral morphology with clear shells in the northwest, and southern parts of the remnant. This morphology could be the effect of a non-uniform ISM, or of a non-uniform ambient magnetic field (Orlando et al. 2007). The remnant has an angular size of $\sim$ 8$\times$8 arcmin, which for a distance of 9.7 kpc (Caswell et al. 1975b) corresponds to a diameter of $\sim$ 20 pc. The MOST radio flux density  at 843 MHz is $S_{\rm 843 MHz}$=18 Jy, which combined with that at 408 MHz $S_{\rm 408 MHz}$=29.8 Jy (Shaver \& Goss 1970) gives a spectral index $\alpha \sim -0.49$ (S $\propto \nu^{\alpha}$).

At the mid-infrared part of the spectrum (Fig.1, central panel), we note a bright filamentary structure running along the north-south direction on the western side of G304.6$+$0.1, which correlates very well with the NW radio shell. This feature, first noted by Reach et al.(2006), is centered at $(\alpha, \delta)_{\rm J2000.0} = (13^{\rm h}05^{\rm m} 30\fs0, -62\degr 41\arcmin 09\arcsec)$ and has an angular size of $\sim$ 6 arcmin. The extended mid-infrared fluxes of this structure at 3.6, 4.5, 5.8, and 8$\mu$m, after the subtraction of intense point-like sources in the region, are: $F_{\rm 3.6 \mu m} \sim$ 1.4 Jy, $F_{\rm 4.5 \mu m} \sim$ 1.46 Jy, $F_{\rm 5.8 \mu m} \sim$ 7.7 Jy, $F_{\rm 8.0 \mu m} \sim$ 21.6 Jy, respectively. Evidently  the Western radio peak agrees very well with the central part of the main mid-infrared arc-like structure. Another remarkable set of small and thin like-arc structure is also clearly noticeable just at the inner boundary of the NW radio shell. All these features are discernible in the four IRAC channels. 

On the contrary, the diffuse X-ray emission observed by the XMM-{\it Newton} telescope (plotted in blue) is characterized by an irregular centrally filled  region, which seems to follow the internal shell-like radio/mid-infrared morphology (see Fig.1, right panel). Another distinct component of the X-ray emission is a faint protruding structure that is clearly revealed at the SW side of the radio shell. This X-ray feature is coincident with a radio feature with a similar morphology, which probably originated in a relatively low-density region of the ISM. Finally, note that within this region  the bright infrared source IRAS 13024-6227 is embedded (Kleinmann et al. 1986), located at $(\alpha, \delta)_{\rm J2000.0} = (13^{\rm h}05^{\rm m} 34\fs0, -62\degr 43\arcmin 35\arcsec)$.
 
\section{Discussion} \label{discusion}

The main radio and X-ray characteristics of G304.6$+$0.1 are a centrally filled X-ray morphology dominated by thermal X-ray emission (92 $\%$ of the total X-ray flux), which is not well correlated with the non-thermal radio shell. This suggests that the SNR belongs to the class of mixed-morphology (MM) remnants (Rho \& Petre 1996). The non-thermal emission represents only a small fraction of the X-ray emission, indicating that only a minor part of the flux could be associated with highly energetic electrons accelerated at the SNR front shock. The detection of OH masers at 1720 MHz by Frail et al.(1996) is an indicator of shock interaction with a molecular cloud, a result that supports the notion that G304.6$+$0.1 is a member of the proposed group of MM SNRs. A list containing SNRs with these characteristics can be found in Rho \& Petre (1998). 

There are about 20 SNRs with convincing evidence for interaction with ambient molecular clouds. A substantial fraction of them are MM or thermal composite SNRs, with central filled thermal X-ray emission surrounded by radio and (in some cases) infrared shell structures (see e.g., Yusef-Zadeh et al. 2003; Yang \& Yang 2007). When a SN begins, the energetic impact of the explosion can drive an asymmetric expanding bubble. If the surrounding ambient gas is non-uniform, the shock-front could severely affect the dynamic structure of the gas and trigger the irregular radio and infrared structures. From Fig.1 (right panel) it is evident that there are many similarities between the radio and mid-infrared morphologies observed on the NW part of G304.6$+$0.1. Additionally, the X-ray emission is anticorrelated with both mid-infrared and radio brightness. 
These results suggest that Kes 17 is propagating on a non-uniform environment and that the shock front is interacting with adjacent massive molecular clouds. The shell-like structures may be caused by shock compression. 

At least four possible scenarios have been introduced in past years to explain thermal X-ray radiation inside
radio shells of SNRs: i) cloudlet evaporation in the SNR interior (White \& Long 1991), ii) thermal conduction smoothing out
the temperature gradient across the SNR and enhancing the central density (Cox et al. 1999), iii) a radiatively cooled rim with a
hot interior (Harrus et al. 1997), and iv) possible collisions with molecular clouds (Safi-Harb et al. 2005).

The source G304.6$+$0.1 seems to propagate into a region of greatly varying density, possibly close to several molecular clouds. The detection of 1720 MHz OH maser emission (see Frail et al. 1996; McDonnell et al. 2008) near G304.6$+$0.1 is an indicator of the interaction with a molecular cloud. Then  the most suitable scenario for describing the SNR evolution is that developed by White \& Long (1991). In these remnants, the clouds were overrun by the shock and are now cooling and evaporating by saturated conduction in the hot post-shock region (Cowie \& McKee, 1977). For this reason, the observed X-ray morphology may be very different than that of ordinary shell-like SNRs.

It is possible to compute some physical parameters of G304.6$+$0.1, using the radio/mid-infrared and X-ray information gathered here. From the X-ray 
image, we can estimate the volume $V$ of the X-ray emitting plasma for the differents regions. Assuming that the plasma fills regions like those observed in Fig. 1 (circles) and assuming a distance of 9.7 kpc, we obtain volumes $V_{\rm north}$= 9.5$\times$10$^{57}$ cm$^{3}$, $V_{\rm center}$= 1.5$\times$10$^{58}$ cm$^{3}$, and $V_{\rm south}$= 1.1$\times$10$^{57}$ cm$^{3}$, respectively. Based on the emission measure (EM) determined by the spectral fitting (see Table 1), we can estimate the electron density of the plasma, $n_{e}$, by $n_{e}$=$\sqrt{EM/V}$, which results in  $n_{e}$(north)$\sim$ 0.99 cm$^{-3}$, $n_{e}$(centre)$\sim$ 0.89 cm$^{-3}$, and $n_{e}$(south)$\sim$ 2.26 cm$^{-3}$, respectively. In this case, the number density of the nucleons was simply assumed to be the same as that of electrons. The age $t$ was then determined from the ionization timescale, $\tau$, by $t$=$\tau$/$n_{e}$. Therefore, the elapsed time after the plasma was heated is in the range of (2.8-6.4)$\times$10$^{4}$ yr. This result shows that G304.6$+$0.1 is a middle-aged SNR. The total mass of the plasma $M_{\rm total}$ was estimated by $M_{\rm total}$=$n_{e}$$V$\,$m_{\rm H}$ $\sim$ 8-15 $M_{\odot}$, where $m_{\rm H}$ is the mass of a hydrogen atom.

Finally, we compared the X-ray luminosity of G304.6$+$0.1 with other prototypical MM SNRs. The unabsorbed X-ray luminosity in the 0.5-8.0 keV band of G304.6$+$0.1 for a distance of 9.7 kpc is $L_{\rm X} \sim 7.5 \times 10^{35}$ erg s$^{-1}$. This value is similar to the SNR G359.1-0.5 (Yusef-Zadeh et al. 1995), but one order of magnitude higher than that of W28 (Claussen et al. 1997), which was computed at the same energy range. 

\section{Conclusions}

We presented the first detailed multiwavelength study of the SNR G304.6$+$0.1, using radio, mid-infrared, and X-ray observations. The X-ray morphology reveals diffuse interior X-ray emission with an irregular filled-center and possible arc-type structure, which seems to follow the internal part of the NW radio shell. The XMM-{\it Newton} EPIC spectrum displays several emission lines, which support the scenario of a thermal nature for the X-ray emission. However, a single-component model fails to describe the global X-ray spectrum, and at least two components are required. The spectrum is well represented by a combination of a thermal and non-thermal models. The mid-infrared observations show a bright filamentary structure along the north-south direction coincident with the NW radio shell. This suggests that Kes 17 is propagating on a nonuniform environment with high density and the front shock is interacting with adjacent massive molecular clouds. The good correspondence of radio and mid-infrared emissions implies that these shells are caused by shock compression. 


The radio, mid-infrared, and X-ray characteristics of G304.6$+$0.1 (centrally filled X-ray morphology dominated by thermal X-ray emission, not correlated with the non-thermal radio shell) indicate that the object is a middle-aged SNR, and that it belongs to the class of mixed-morphology remnants. 
Future Chandra observations could help to reveal more details of the X-ray structure and to discern the probable presence of a compact point-like X-ray source inside the remnant. 

\begin{acknowledgements}
      
We are grateful to the referee for her/his valuable suggestions and comments, which helped us to improve the paper. The authors acknowledge support by DGI of the Spanish Ministerio de Educaci\'on y Ciencia under grants AYA2007-68034-C03-02/-01, FEDER funds, Plan Andaluz de Investigaci\'on Desarrollo e Innovaci\'on (PAIDI) of Junta de Andaluc\'{\i}a as research group FQM-322 and the excellence fund FQM-5418. J.A.C., J.F.A.C. and G.E.R are researchers of CONICET. J.F.A.C was suportes by grant PICT 2007-02177 (SecyT). G.E.R. and J.A.C. were supported by grant PICT 07-00848 BID 1728/OC-AR (ANPCyT) and PIP 2010-0078 (CONICET). J.L.S. acknowledges support by the Spanish Ministerio de Innovaci\'on y Tecnolog\'ia under grant AYA2008-06423-C03-03.  
          
\end{acknowledgements}

\end{document}